\documentclass[journal,comsoc]{IEEEtran}
\usepackage[T1]{fontenc}

\usepackage{cite}

\ifCLASSINFOpdf
\else
\fi
\usepackage{amsmath}
\usepackage[cmintegrals]{newtxmath}
\usepackage{xcolor}

\usepackage{soul}
\usepackage{url}
\usepackage{hyperref}

\usepackage{longtable}
\usepackage{multirow}
\usepackage{tabularx}
    \newcolumntype{L}{>{\raggedright\arraybackslash}X}
\usepackage{booktabs}
\usepackage{subcaption}

\usepackage[ruled,linesnumbered,vlined]{algorithm2e}
\usepackage{caption}
\usepackage{graphicx}
\usepackage{utfsym}

\usepackage[flushleft]{threeparttable}
\SetKwRepeat{Do}{do}{while}

\usepackage{xcolor}
\usepackage{bm}
\begin{document}
%
\title{
Collaborative Zone-Adaptive Zero-Day Intrusion Detection for IoBT
}


\author{\IEEEauthorblockN{Amirmohammad Pasdar, Shabnam Kasra Kermanshahi, Nour Moustafa, Van-Thuan Pham}\\
\thanks{Amirmohammad Pasdar and Van-Thuan Pham are with the School of Computing and Information Systems, The University of Melbourne, VIC 3010, Australia (e-mail: apasdar@unimelb.edu.au, thuan.pham@unimelb.edu.au). (\textit{Corresponding author: Amirmohammad Pasdar.})}
\thanks{Nour Moustafa and Shabnam Kasra Kermanshahi are with the School of Systems \& Computing, University of New South Wales, ACT 2612, Australia (e-mail: nour.moustafa@unsw.edu.au; s.kasra\_kermanshahi@unsw.edu.au).}

}


\maketitle


\begin{abstract}
The Internet of Battlefield Things (IoBT) relies on heterogeneous, bandwidth-constrained, and intermittently connected tactical networks that face rapidly evolving cyber threats. In this setting, intrusion detection cannot depend on continuous central collection of raw traffic due to disrupted links, latency, operational security limits, and non-IID traffic across zones. We present \textbf{Z}one-\textbf{A}daptive \textbf{I}ntrusion \textbf{D}etection (ZAID), a collaborative detection and model-improvement framework for \textit{unseen attack types}, where ``zero-day'' refers to previously unobserved attack families and behaviours (not vulnerability disclosure timing). ZAID combines a universal convolutional model for generalisable traffic representations, an autoencoder-based reconstruction signal as an auxiliary anomaly score, and lightweight adapter modules for parameter-efficient zone adaptation. To support cross-zone generalisation under constrained connectivity, ZAID uses federated aggregation and pseudo-labelling to leverage locally observed, weakly labelled behaviours. We evaluate ZAID on ToN\_IoT using a zero-day protocol that excludes MITM, DDoS, and DoS from supervised training and introduces them during zone-level deployment and adaptation. ZAID achieves up to 83.16\% accuracy on unseen attack traffic and transfers to UNSW-NB15 under the same procedure, with a best accuracy of 71.64\%. These results indicate that parameter-efficient, zone-personalised collaboration can improve the detection of previously unseen attacks in contested IoBT environments.
\end{abstract}


%
\IEEEpeerreviewmaketitle

\section{Introduction}
With the rapid expansion of the Internet of Things (IoT), the world is experiencing an explosion of connected devices and their use across diverse domains \cite{qadri2020future, xu2023survey}, leading to large-scale data traffic over different networks and protocols. A similar trend is increasingly apparent in modern military environments, often referred to as the Internet of Battlefield Things (IoBT), where connected sensing, communications, and autonomous systems are integrated to improve situational awareness and coordinated operations \cite{strayer2022military, michalski2019internet, wigness2022internet}.

IoBT connects tactical sensors, unmanned platforms, radios, and mission systems in contested environments where connectivity is intermittent, bandwidth is limited, and adversaries can disrupt infrastructure or compromise nodes \cite{li2020research, agadakos2019resilient}. These limitations render purely centralised security monitoring fragile, as routing all traffic to a central analysis node can cause unacceptable latency, create single points of failure, and increase the risk of sensitive operational data exposure. In practice, defensive capabilities must be deployed closer to tactical zones while still leveraging collective learning across the force.

Intrusion detection in IoBT must therefore address three realities simultaneously. First, \emph{heterogeneity and non-IID traffic} as different tactical zones (e.g., platoons, vehicles, command posts) display distinct traffic patterns, assets, and attack surfaces. Second, \emph{domain shift and rapid evolution} implies adversaries can introduce new behaviours and previously unseen attack variants, creating zero-day scenarios in which attack families or behaviours are absent from available training data. Third, \emph{limited labelling and constrained resources}, in which accurate labels are costly in operational contexts, while edge gateways have less compute and memory compared to data-centre deployments. These factors challenge traditional learning-based intrusion detection systems that assume stable distributions, abundant labels, and centralised data collection.

Federated learning (FL) offers a promising approach by enabling collaborative model training without transmitting raw data across zones, thus supporting privacy-by-design and operational constraints \cite{pmlr-v54-mcmahan17a, kairouz2021advances}. However, applying FL directly to IoT or IoBT intrusion detection remains difficult in the presence of severe non-IID data and intermittent connectivity. This is because local biases can dominate updates, global models may underperform in minority zones, and purely supervised objectives can struggle to adapt when unseen behaviours emerge. Recent surveys emphasise both the potential of FL-based intrusion detection and ongoing challenges, including robustness to distribution shifts, label scarcity, and deployment constraints \cite{agrawal2022flids, belenguer2025reviewflids}.

At the same time, modern network intrusion detection increasingly relies on \emph{flow-level and statistical features} that do not require payload inspection, which is important in environments where traffic may be encrypted and deep packet inspection is infeasible \cite{wang2022machine, sarhan2022featuresets}. Deep neural networks have been used for network anomaly and intrusion detection \cite{mohamed2023digital, bierbrauer2023transfer}, but a common challenge in tactical and edge deployments is how to personalise models to different zones without retraining or sharing raw traffic, while still maintaining robust representations learned from large datasets.

\begin{figure}[!t]
\centering
\includegraphics[width=0.48\textwidth]{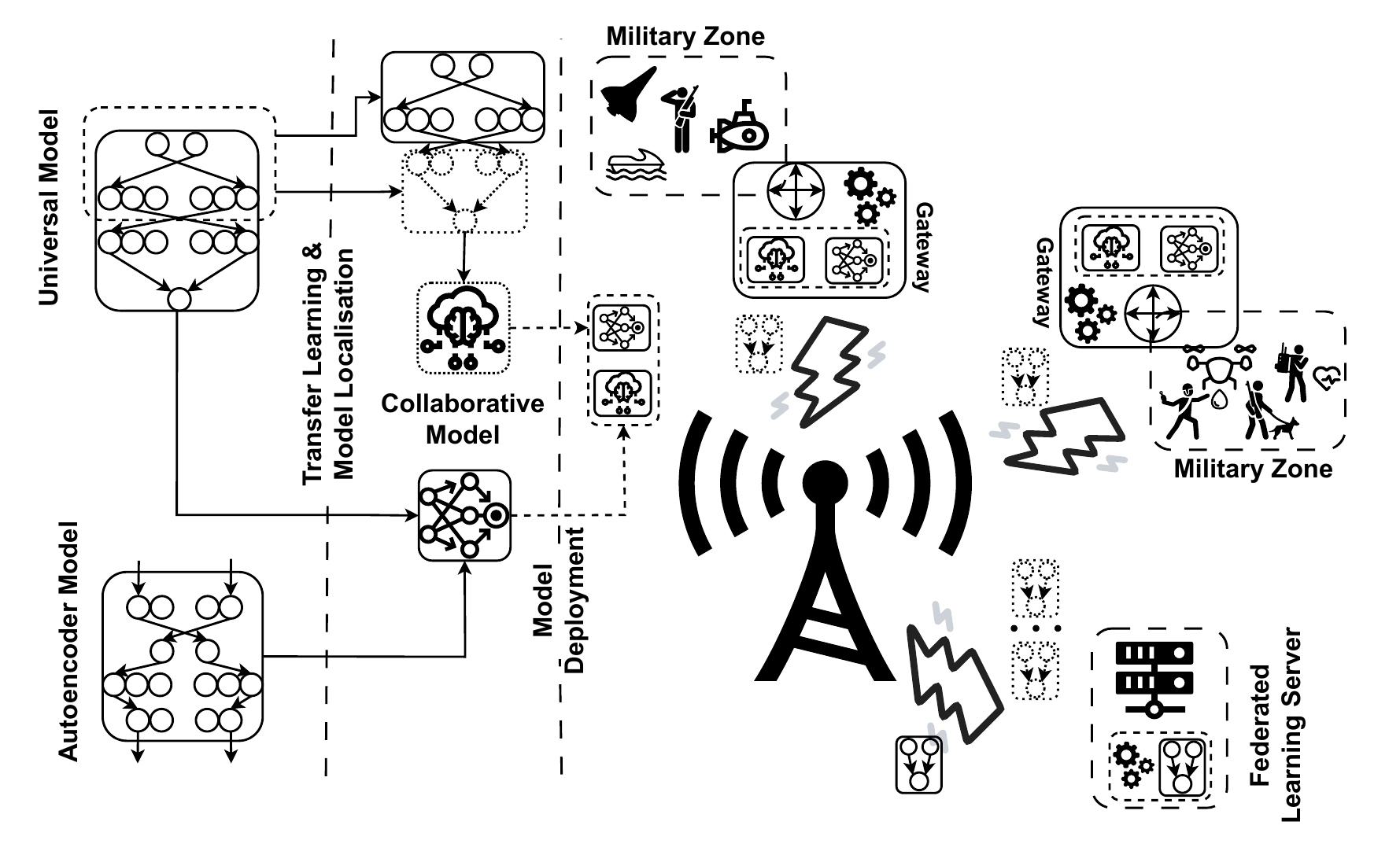}
\caption{Overview of ZAID in an IoBT environment. Each zone processes traffic locally at a restricted gateway, where a universal classifier makes a baseline decision, lightweight adapters customise the classifier to local traffic, and an autoencoder-based anomaly score reliably detects unseen attack types. Zones periodically share only a small set of trainable parameters for aggregation, enabling collaborative enhancement without centralising raw traffic.}
\label{fig:cant_model}
\end{figure}

To address these issues, we propose \textbf{Z}one-\textbf{A}daptive \textbf{I}ntrusion \textbf{D}etection (ZAID), a collaborative, zero-day intrusion-detection approach for IoBT-style environments. ZAID, shown in Figure \ref{fig:cant_model}, is built around a hybrid pipeline where a universal model provides a shared representation of traffic behaviour, and an autoencoder-derived reconstruction signal offers an auxiliary anomaly score for deviations from normal traffic. Importantly, ZAID introduces parameter-efficient zone adaptation through adapter modules, enabling each zone to specialise locally while keeping most of the universal model frozen. This design reduces the amount of trainable state that needs to be updated and communicated, and it supports collaborative improvement via federated aggregation without transferring raw network traffic.

The contributions of this paper are:
\begin{itemize}
    \item We formulate a collaborative intrusion detection setting for IoBT where traffic is partitioned across tactical zones with heterogeneous distributions and intermittent connectivity, motivating collaboration without centralised raw traffic collection.
    \item We design ZAID as a hybrid collaborative pipeline that combines a universal convolutional model, an autoencoder-derived anomaly signal, and federated collaboration to detect previously unseen attack behaviours under non-IID data.
    \item We introduce complementary adaptation mechanisms, such as federated collaboration, adapter-based specialisation, and pseudo-labelling, to improve cross-zone generalisation while keeping gateway-side updates lightweight.
    \item We evaluate ZAID using ToN\_IoT and UNSW-NB15 under a zero-day protocol consistent with the current implementation, and we analyse cross-zone and cross-domain behaviour.
\end{itemize}

The remainder of this paper is organised as follows: Section~\ref{sec:rw} reviews related work; Section~\ref{sec:cad} details the ZAID design; Section~\ref{sec:eval} presents evaluation results and discusses limitations and future directions; and Section~\ref{sec:conclusion} concludes. 

\section{Related Work}
\label{sec:rw}

IoBT security monitoring shares many of the same technical attack surfaces as civilian IoT, but it operates under significantly different constraints, such as intermittent connectivity, limited bandwidth, mission-driven prioritisation, and active adversarial disruption of infrastructure and endpoints \cite{wigness2022internet, agadakos2019resilient, li2020research, strayer2022military}. These constraints challenge traditional intrusion detection systems that assume (1) stable and centrally observable traffic, (2) abundant labelled attack traces, and (3) consistent traffic patterns across deployment sites. In the following, we review the closest research areas that inform ZAID, including edge/gateway intrusion detection systems (IDS) under constraints, collaborative/federated intrusion and anomaly detection, and transfer/personalisation mechanisms for non-IID and weakly labelled settings.

\subsection{Intrusion detection at IoT/IoBT gateways under constrained operation}
A substantial amount of research focuses on learning-based intrusion detection for IoT traffic, including deep models and hybrid methods that combine feature engineering and selection with supervised classification \cite{survey19, survey11, vasilomanolakis2015taxonomy}. Practical deployments often place IDS functionality at gateways or edge nodes to avoid continuously streaming raw traffic to a central site, thereby reducing bandwidth consumption and exposure of sensitive operational data. Representative work includes adaptive IDS designs (e.g., ensembles and time-varying feature selection) \cite{rp12, rp27} and IoT-specific detection studies on common attack classes such as DDoS \cite{rp67}. A recurring concern is that reported performance can be sensitive to the chosen feature set and dataset artefacts; recent work highlights the importance of evaluating whether a ``standard'' feature set generalises across realistic distributions and deployments \cite{sarhan2022featuresets}. These observations highlight ZAID's focus on general, protocol-agnostic traffic features that remain observable even when payloads are encrypted, as well as the importance of explicit evaluation under distribution shift across zones.

\subsection{Collaborative and federated intrusion/anomaly detection}
Collaborative intrusion detection systems (CIDS) have long aimed to share security knowledge across monitors while limiting the need to transfer raw telemetry \cite{vasilomanolakis2015taxonomy}. In tactical and mobile network contexts, decentralised collaboration is often proposed to reduce single points of failure and align with distributed network architectures (e.g., in highly dynamic vehicular settings) \cite{rp35}. Federated learning provides a modern mechanism to support such collaboration by aggregating model updates rather than raw data \cite{pmlr-v54-mcmahan17a, kairouz2021advances}. Within IoT security, several systems apply FL to anomaly/IDS objectives, such as DIoT, which uses federated aggregation of device-type behavioural models and argues that FL can support emerging/unknown attacks without centralising traffic \cite{nguyen2019diot}; Mothukuri \emph{et al.} propose FL-based anomaly detection for IoT security attacks using decentralised data \cite{mothukuri2022flad}; Popoola \emph{et al.} present federated deep learning for zero-day botnet detection at the edge \cite{popoola2021federated}. Recent surveys consolidate these directions and emphasise that severe non-IID data, limited labels, and adversarial manipulation remain core unresolved obstacles for FL-based IDS in the wild \cite{agrawal2022flids, belenguer2022reviewflids}. Beyond IDS-specific FL, adjacent FL-based security frameworks have also been explored under neighbouring problem framings, including federated learning for IoT digital forensics and federated defence designs for industrial/supply-chain network settings \cite{mohamed2023digital, rp11}. These are treated as contextual reference points, not protocol baselines, but both aim to maintain telemetry locality and enable collaborative learning across distributed participants.

\subsection{Transfer, personalisation, and weak supervision under non-IID shift}
Cross-domain and cross-site IDS has repeatedly shown that distribution shift across networks, i.e., protocol mixes, topologies, device populations, and operational phases, can break naive generalisation, motivating transfer learning and domain adaptation \cite{bierbrauer2023transfer}. In FL, non-IID client data can induce ``client drift'' and degrade global models; algorithmic mitigations such as FedProx and SCAFFOLD explicitly target this issue \cite{li2020fedprox, karimireddy2020scaffold}, while personalised FL methods optimise a global model alongside local personalisation \cite{pmlr-v139-li21h}. Separately, parameter-efficient transfer methods (e.g., adapter modules) have been widely used in other domains to specialise models with small trainable components while largely preserving a shared backbone \cite{pmlr-v97-houlsby19a}. Semi-supervised learning via pseudo-labelling is also a standard technique to exploit weakly labelled local samples \cite{lee2013pseudo}. ZAID is informed by these strands; hence, rather than assuming fully supervised labels per zone, ZAID explicitly supports zone-level specialisation with lightweight adapters and pseudo-labelling on top of collaborative FL updates, aiming to improve robustness under tactical non-IID shifts.

\subsection{Threats to collaborative IDS: adversarial robustness}
Finally, collaborative and FL-based IDS introduce additional attack surfaces, e.g., poisoning of updates or data-driven manipulation that can degrade detection. Recent work demonstrates the practicality of poisoning and evasion-style threats against network intrusion detection pipelines \cite{mohammadian2024poisoning}. This motivates ZAID's explicit positioning on threat assumptions and treating robustness under adversarial participation as a key limitation and future work item.

Therefore, ZAID targets the intersection of these themes, first, IoBT operational constraints that discourage continuous centralised raw traffic collection \cite{wigness2022internet, strayer2022military}, second, zero-day conditions under domain shift and scarce labels \cite{bierbrauer2023transfer, sarhan2022featuresets}, and third, the need for collaboration across zones without collapsing under non-IID heterogeneity \cite{kairouz2021advances, agrawal2022flids}. ZAID differs from prior FL-based IDS efforts by combining (a) a shared universal representation learned collaboratively, (b) an anomaly-oriented component to emphasise previously unseen behaviours, and (c) three complementary adaptation mechanisms (federated collaboration, adapter-based personalisation, and pseudo-labelling) to support zone-level deployment under constrained compute and communications. Table \ref{tbl:rw_summary} provides a high-level comparison between ZAID and representative studies.

\begin{table*}[!t]
\caption{Representative related work and how it connects to ZAID. ``Domain'' is stated explicitly to avoid over-claiming ``IoBT'' when a study is evaluated in civilian IoT or other distributed settings.}
\centering
\footnotesize
\setlength{\tabcolsep}{3pt}
\renewcommand{\arraystretch}{1.15}
\begin{tabularx}{\linewidth}{@{}l p{2.2cm} c p{2.3cm} p{2.5cm} X@{}}
\toprule
Study & Domain & No raw data sharing & Collaboration mechanism & Personalisation /Transfer & Notes \\
\midrule
\cite{rp35} & Vehicular / mobile (IoV) & Yes & Distributed CIDS & Limited & Privacy-preserving collaborative IDS under mobility \\
\cite{nguyen2019diot} & IoT gateways & Yes & Federated aggregation & Device-type models & Unsupervised/self-learning behavioural profiling with FL \\
\cite{mothukuri2022flad} & IoT & Yes & Federated learning & Limited & FL-based anomaly detection on decentralised data \\
\cite{popoola2021federated} & IoT edge & Yes & Federated learning & Limited & Zero-day botnet detection framed in FL setting \\
\cite{bierbrauer2023transfer} & Network IDS (cross-domain) & N/A & N/A & Yes & Highlights distribution shift and transfer considerations \\
\cite{sarhan2022featuresets} & Network IDS (features) & N/A & N/A & Indirect & Evaluates whether standard feature sets generalise \\
\cite{mohamed2023digital} & IoT (forensics/security) & Yes & Federated learning & Limited & FL-based digital forensics in IoT environments (also used as a reference-inspired baseline) \\
\cite{rp11} & Supply chain / IIoT & Yes & Federated defence framework & Limited & Deep federated defence framework for SC4.0 networks (also used as a reference-inspired baseline) \\
\cite{rp12} & IoT gateways & N/A & Standalone (gateway IDS) & N/A & Adaptive IDS at IoT gateways (non-federated) \\
ZAID (this work) & IoBT-style zones (simulated) & Yes & FL + local adaptation & Yes (adapters) & Adds adapters + pseudo-labelling for zone specialisation \\
\bottomrule
\end{tabularx}
\label{tbl:rw_summary}
\end{table*}

\section{\textbf{Z}one-\textbf{A}daptive \textbf{I}ntrusion \textbf{D}etection (\textbf{ZAID})}
\label{sec:cad}

This section formulates the collaborative zero-day detection setting and then details the ZAID components (see Figure~\ref{fig:cant_model} for an overview). We consider ``zero-day'' in the operational sense of \emph{previously unobserved attack families or behaviours} in the deployment zone, rather than the timing of vulnerability disclosures.

\subsection{System model and threat setting}
We consider an IoBT environment partitioned into a set of tactical zones $\mathbb{Z}=\{z_1,\dots,z_{|\mathbb{Z}|}\}$. Each zone $z_i$ contains assets communicating via heterogeneous networks and protocols, generating local network traffic $NC_{z_i}$. Each zone is monitored by a gateway (or edge security device) with limited processing capacity, which extracts flow-level and statistical features, performs local inference, and executes periodic local updates. Using flow/statistical features (rather than payload inspection) is compatible with settings where traffic may be encrypted and deep packet inspection is infeasible, and it is common in practical IDS feature engineering \cite{sarhan2022featuresets, wang2022machine}.

The objective is to detect abnormal traffic and improve robustness against unseen attack behaviours under (1) \emph{non-IID zone distributions} where different device populations, operational phases, and mission roles,
(2) \emph{limited labels}, in which manual labelling is costly in operational settings, and
(3) \emph{constrained connectivity} such that zones may not continuously send raw traffic to a central site.

\subsection{From centralised traffic analysis to collaborative zone learning}
\begin{figure}[!b]
  \centering
    \begin{minipage}[b]{0.36\columnwidth}
    \centering
    \includegraphics[scale=0.30,keepaspectratio]{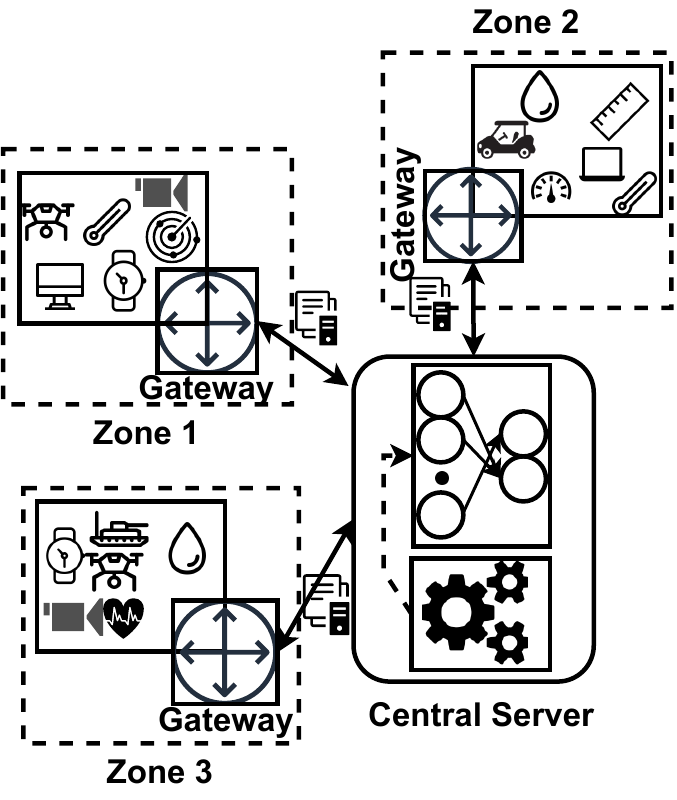}
    \subcaption{Centralised (traffic centralisation baseline)}
    \label{fig:strg_cnna}
  \end{minipage}  
  \begin{minipage}[b]{0.49\columnwidth}
  \centering
    \includegraphics[scale=0.30,keepaspectratio]{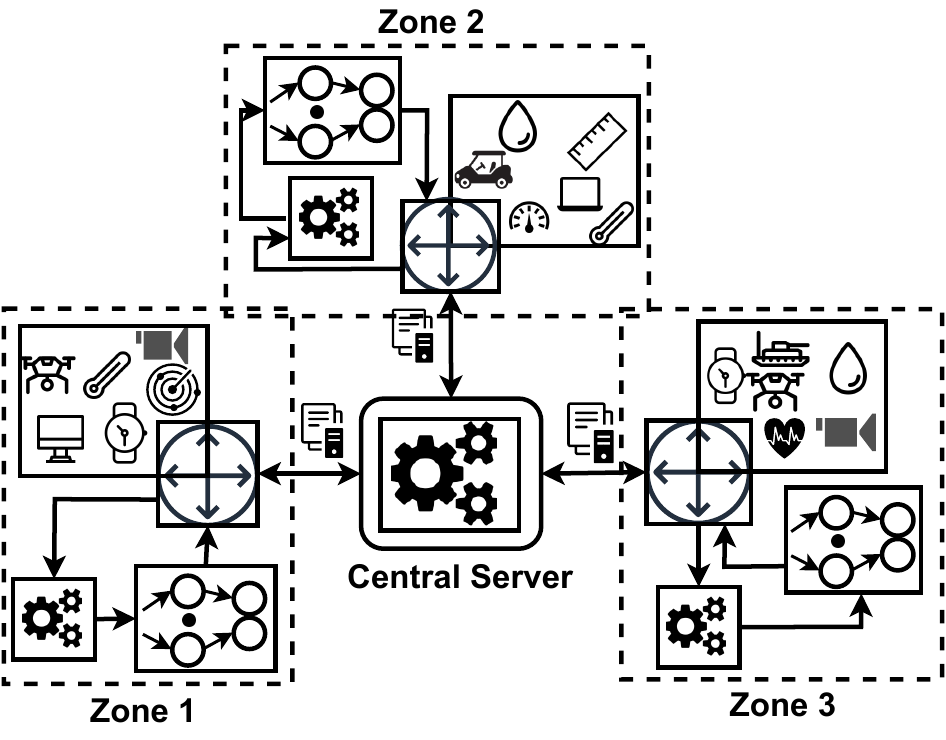}
    \subcaption{Zone-local learning with collaborative updates (ZAID)}
    \label{fig:strg_cnnb}
  \end{minipage}
 \caption{Conceptual contrast between (a) centralised traffic analysis/training and (b) ZAID-style zone-local learning that shares model updates/parameters rather than raw traffic.}
 \label{fig:strg_cnn}
\end{figure}

A conventional baseline centralises telemetry by routing traffic to a central analysis node for preprocessing, training, and detection. In contested or bandwidth-limited environments, this can be operationally fragile. ZAID instead emphasises \emph{zone-local detection and learning}, while enabling collaboration by exchanging \emph{model updates and lightweight trainable parameters}, rather than raw traffic. Figure~\ref{fig:strg_cnn} illustrates the conceptual contrast between (a) centralised traffic analysis and (b) ZAID-style zone learning with collaborative parameter exchange. This design reduces the need to transmit raw traffic across the network and avoids concentrating all processing at a single point of failure, while still allowing cross-zone sharing of learned representations \cite{pmlr-v54-mcmahan17a, kairouz2021advances}.

\subsection{ZAID Universal ($\mathcal{U}$) and Autoencoder ($\mathcal{A}$) models}
ZAID uses two base models trained on the available labelled dataset: a universal convolutional classifier $\mathcal{U}$ and an autoencoder $\mathcal{A}$.

\begin{figure}[!b]
    \centering
\begin{minipage}[b]{\columnwidth}
    \centering    \includegraphics[keepaspectratio,scale=0.42]{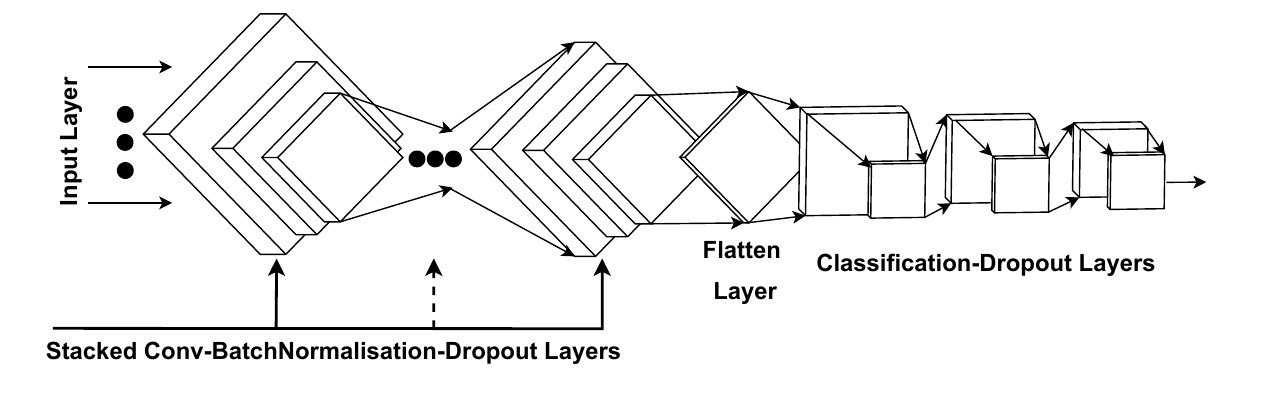}
    \subcaption{ZAID universal model with three stacked convolutional layers with filters 64, 32, 16, ReLU activation and kernel size 3, followed by 256, 64, 16 dense layers with 0.2 for dropout.}
    \label{fig:prc_layer_cnna}
    \end{minipage}

    \begin{minipage}[b]{\columnwidth}
    \centering    \includegraphics[keepaspectratio,scale=0.48]{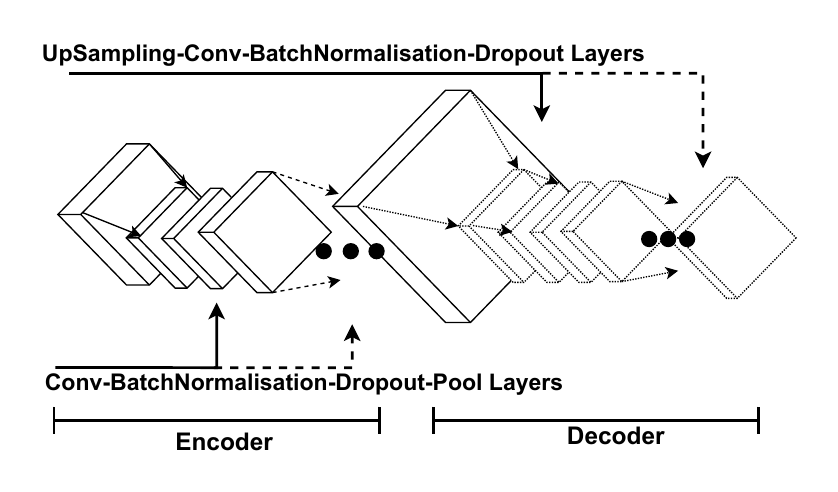}
    \subcaption{ZAID autoencoder structure with two stacked convolutional-pooling layers with filters 16, 32, a bottleneck layer with 64 filters, and kernel size 3, ending at two stacked upsampling-convolutional layers, with 0.2 for dropout.}
    \label{fig:prc_layer_cnnb}
    \end{minipage}
    \caption{Architectures used in the current ZAID implementation: (a) universal CNN classifier $\mathcal{U}$ and (b) autoencoder $\mathcal{A}$ for reconstruction-based anomaly signal.}
    \label{fig:prc_layer_cnn}
\end{figure}

\textbf{Universal model $\mathcal{U}$.} The universal model is a multi-layer CNN trained for binary classification, i.e., normal vs abnormal. The intent is to learn a generalisable representation that can be reused across zones as an initial model for deployment and further local adaptation. Figure~\ref{fig:prc_layer_cnna} depicts the architecture used in the current implementation, consisting of stacked convolutional blocks (Conv + BatchNorm + Dropout) followed by fully connected layers and a sigmoid output for binary classification. The CNN outputs a probability $\lambda_{\mathcal{U}} \in [0,1]$ indicating the likelihood of abnormal traffic.

\textbf{Autoencoder $\mathcal{A}$.} ZAID also uses an autoencoder trained on \emph{normal} (or normal-dominant) traffic to learn reconstruction of expected behaviour. Figure~\ref{fig:prc_layer_cnnb} shows a convolutional encoder--decoder structure with a bottleneck representation and symmetric upsampling for reconstruction. Given an input feature vector (or sequence) $x$, the autoencoder produces a reconstruction $\hat{x}$ and a reconstruction error (e.g., $\|x-\hat{x}\|$). We denote the normalised reconstruction-based anomaly signal by $\lambda_{\mathcal{A}} \in [0,1]$, in which higher indicates stronger deviation from learned normal patterns. This signal provides an auxiliary anomaly-oriented indicator that can be useful when a zone encounters behaviours not well covered by supervised training data.

\subsection{Feature selection and training of $\mathcal{U}$ and $\mathcal{A}$}
To reduce dimensionality and focus learning on informative characteristics, ZAID applies feature selection before training. The current pipeline uses a Random Forest Classifier (RFC) with Recursive Feature Elimination (RFE) to obtain a stability-ranked feature list across stratified splits, followed by Mutual Information (MI) scoring to retain highly informative features. This selection is then used for training both $\mathcal{U}$ and $\mathcal{A}$.

\begin{algorithm}[!b]
\SetEndCharOfAlgoLine{}
\setcounter{AlgoLine}{0}
\SetKwFunction{procr}{$FeatureSelection$}
  \KwData{Network traffic dataset $(X_{NC},Y_{NC})$, number of splits $n$, test size $m$, quantile threshold $r$.}
 \KwResult{Selected feature set $\mathcal{F}_t$.}
  \SetKwProg{Proc}{Procedure}{}{}
  \Proc{\procr{$X_{NC}, Y_{NC}$}}{
  Preprocess $NC$\;
  $ranking \gets []$\;
  $\mathcal{S} \gets$ StratifiedShuffleSplit$(n,m)$\;
  \For{$i \leq n$}{
        $rfc \gets$ Initialise Random Forest Classifier\;
        $\_rnk \gets$ RFE$(\mathcal{S}_i, rfc)$\;
        $ranking.append(\_rnk)$\;
  }
  $rankings \gets$ Sort(Mean($ranking$), ascending)\;
  $\mathcal{F}^{MI}_{score} \gets MI(rankings)$\;
  \nl \KwRet $\mathcal{F}_t \gets$ Quantile$(\mathcal{F}^{MI}_{score}, r)$\;
  }
\SetKwFunction{procr}{$\mathcal{U}$-Training}
  \KwData{Dataset $(X_{NC},Y_{NC})$, epochs $e$, selected features $\mathcal{F}_t$.}
 \KwResult{Universal model $\mathcal{U}$ with weights $\bm{\theta}_{\mathcal{U}}$.}
  \Proc{\procr{$X_{NC},Y_{NC},\mathcal{F}_t$}}{
  Initialise $\bm{\theta}_{\mathcal{U}}$\;
  $X_{NC} \gets$ MinMaxScaler$(X_{NC}(\mathcal{F}_t))$\;
  \For{$i \leq e$}{
        \For{$\forall (x,y) \in (X_{NC},Y_{NC})$}{
            $y' \gets CNN_{\triangleright}(x;\bm{\theta}_{\mathcal{U}})$\;
            $CNN_{\triangleleft}(E(y,y');\bm{\theta}_{\mathcal{U}})$\;
        }
  }
  \nl \KwRet $\mathcal{U}, \bm{\theta}_{\mathcal{U}}$\;
  }
  \SetKwFunction{procr}{$\mathcal{A}$-Training}
  \KwData{Normal(-dominant) dataset $X^{*}_{NC}$, epochs $e$, selected features $\mathcal{F}_t$.}
 \KwResult{Autoencoder $\mathcal{A}$ with weights $\bm{\theta}_{\mathcal{A}}$.}
  \Proc{\procr{$X^{*}_{NC},\mathcal{F}_t$}}{
  Initialise $\bm{\theta}_{\mathcal{A}}$\;
  $X^{*}_{NC} \gets$ MinMaxScaler$(X^{*}_{NC}(\mathcal{F}_t))$\;
  \For{$i \leq e$}{
        \For{$\forall x \in X^{*}_{NC}$}{
            $x' \gets AE_{\triangleright}(x;\bm{\theta}_{\mathcal{A}})$\;
            $AE_{\triangleleft}(E(x,x');\bm{\theta}_{\mathcal{A}})$\;
        }
  }
  \nl \KwRet $\mathcal{A}, \bm{\theta}_{\mathcal{A}}$\;
  }
\caption{Training pipeline for universal model $\mathcal{U}$ and autoencoder $\mathcal{A}$.}
\label{algo:cnn_trn}
\end{algorithm}

Algorithm~\ref{algo:cnn_trn} summarises the training pipeline. Feature selection returns $\mathcal{F}_t$, which is then used to train the universal classifier $\mathcal{U}$ and the autoencoder $\mathcal{A}$. We describe the optimisation abstractly using forward/backward passes; the concrete optimiser and loss function follow the implementation settings used in the evaluation section.

\subsection{ZAID localised and collaborative model ($\mathcal{C}$)}
\label{subsec:cad_collab}

\begin{figure}[!b]
    \centering
    \begin{minipage}[b]{\columnwidth}
    \centering \includegraphics[keepaspectratio,scale=0.42]{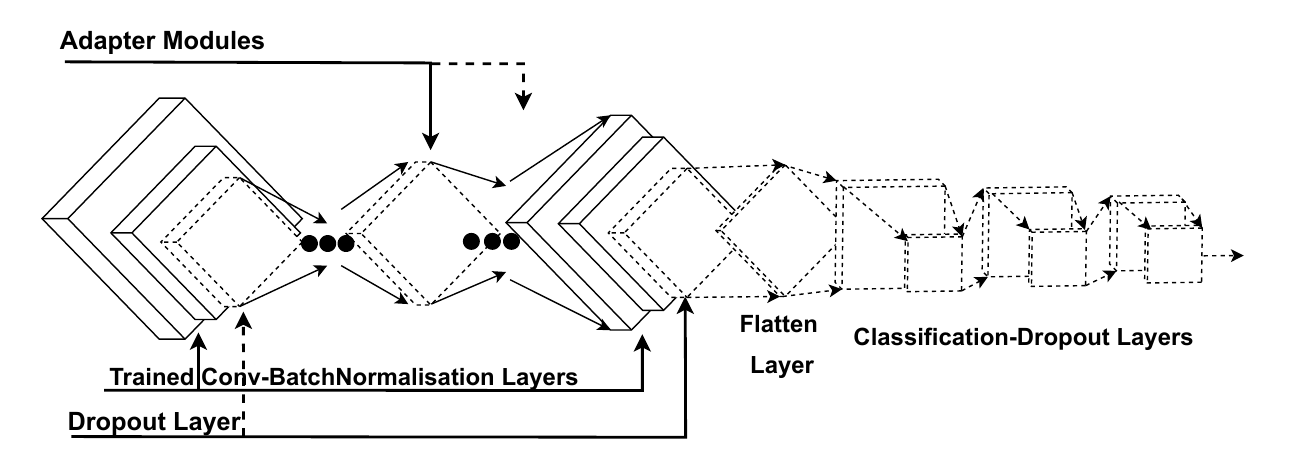}
    \caption{Zone localisation: the universal backbone is reused while lightweight adapter modules and classification layers are trainable (dashed blocks).}
    \label{fig:localized_dnna}
    \end{minipage}
\end{figure}

\textbf{Zone localisation via adapters.}
ZAID deploys a zone-local model $\mathcal{C}_{z_i}$ by attaching lightweight adapter modules to the universal backbone. Adapters are small trainable modules designed for parameter-efficient specialisation, while keeping most of the base model frozen \cite{pmlr-v97-houlsby19a, rebuffi2017residualadapters}. In the current design (Figure~\ref{fig:localized_dnna}), adapters follow a bottleneck pattern implemented using $1\times1$ convolutions (down-projection, non-linearity, up-projection) with residual connections. This enables zone-specific adjustments without retraining the full universal model.

\textbf{Pseudo-labelling for weakly labelled traffic.}
In deployment, some zone traffic may be unlabelled. ZAID uses pseudo-labelling \cite{lee2013pseudo} to assign approximate labels when model confidence is sufficiently high, supporting semi-supervised updates of $\mathcal{C}_{z_i}$. The autoencoder anomaly signal serves as an auxiliary indicator of deviations from expected normal behaviour, enabling flagging of uncertain cases for further inspection rather than assigning a confident label.

\textbf{Weighted anomaly score for zone decision-making.}
To integrate the universal classifier, zone-local classifier, and autoencoder anomaly signal, ZAID defines a weighted score:
\begin{equation}
\label{eq:anomaly_score}
AS_{\omega} =
\omega_{\alpha} \cdot \lambda_{\mathcal{A}} +
\omega_{\gamma} \cdot \lambda_{\mathcal{C}} +
\omega_{\beta} \cdot \lambda_{\mathcal{U}},
\end{equation}
where $\lambda_{\mathcal{U}}$ is the universal model probability of abnormal traffic, $\lambda_{\mathcal{C}}$ is the zone-local model probability of abnormal traffic, and $\lambda_{\mathcal{A}}$ is the normalised reconstruction-based anomaly signal. The weights $\omega_{\alpha},\omega_{\beta},\omega_{\gamma}$ control the relative influence of the three components, and their selection follows the implementation used in evaluation. This formulation allows ZAID to incorporate an anomaly-oriented signal alongside supervised predictions when operating under unseen behaviours.

\subsection{Federated collaboration and aggregation}
ZAID supports collaborative improvement across zones by exchanging model updates rather than raw traffic. Let $\mathcal{D}_{z_i}=\{(x^i_j,y^i_j)\}_{j=1}^{m_i}$ denote the (labelled and pseudo-labelled) training set available at zone $z_i$, where $x^i_j \in \mathbb{R}^{d}$ is the feature vector and $y^i_j \in \{0,1\}$ is the associated label. Let $\bm{\theta}$ denote the parameters of the shared model state distributed to zones. In ZAID, this can include the universal backbone state and/or collaborative components, while adapter parameters remain zone-specific.

Federated learning aims to find parameters $\bm{\theta}$ that minimise the weighted empirical loss across zones:
\begin{equation}
\min_{\bm{\theta}} \sum_{i=1}^{|\mathbb{Z}|} \frac{m_i}{\sum_{k=1}^{|\mathbb{Z}|} m_k}
\sum_{j=1}^{m_i} \ell\big(f(x^i_j;\bm{\theta}), y^i_j\big),
\label{eq:flagg}
\end{equation}
where $\ell(\cdot)$ is a loss function and $f(\cdot;\bm{\theta})$ is the model. In FedAvg \cite{pmlr-v54-mcmahan17a}, at each communication round $t$, the coordinator sends the current parameters $\bm{\theta}_t$ to participating zones. Each zone performs local optimisation to obtain updated parameters $\bm{\theta}^{i}_t$ and returns them to the coordinator, which aggregates using a weighted average:
\begin{equation}
\bm{\theta}_{t+1} =
\sum_{i=1}^{|\mathbb{Z}|} \frac{m_i}{\sum_{k=1}^{|\mathbb{Z}|} m_k}\bm{\theta}^{i}_t.
\label{eq:fedavg}
\end{equation}
This collaborative procedure enables learning from distributed and potentially non-IID data while keeping raw traffic local to each zone \cite{kairouz2021advances}. In ZAID, this collaborative state complements zone-local adapters, which are trained locally to personalise detection without requiring full model fine-tuning.

\section{Evaluation}
\label{sec:eval}
This section evaluates ZAID under an IoBT-motivated deployment in which traffic is partitioned across zones, labels are limited, and previously unseen attack families may emerge after deployment. We first describe the datasets, zero-day protocol, and metrics. We then evaluate (1) the universal model and autoencoder, (2) collaborative deployment preparation and pseudo-labelling, (3) collaborative improvement via federated aggregation, and (4) cross-domain transfer to UNSW-NB15.

\subsection{Datasets and experimental protocol}
\label{sec:eval_setup}
\textbf{ToN\_IoT.} We use the ToN\_IoT dataset \cite{alsaedi2020toniot} for the main experiments. It contains large-scale IoT/IIoT network traffic with diverse attack scenarios, including scanning/reconnaissance and exploitation behaviours (e.g., injection, DoS/DDoS, and MITM). Out of more than 5.3 million captured traffic records, approximately 50\% of the dataset is used to train the \emph{universal model}, and the remainder is used to simulate zone-level adaptation for localised and collaborative models.

\textbf{Zero-day protocol (withheld attack families).} To approximate zero-day conditions, three minority attack categories are \emph{excluded} from supervised training of the universal model and only appear later during zone-level adaptation and collaboration, such as \textbf{MITM}, \textbf{DDoS}, and \textbf{DoS}. Table~\ref{tbl:attack_ratio_train_toniot} lists the attack proportions in the ToN\_IoT \emph{training} subset after removing these three withheld categories.

\begin{table}[!b]
\footnotesize
\renewcommand\arraystretch{1.1}
\centering
\caption{Attack proportion in training dataset (ToN\_IoT) after withholding MITM, DDoS, and DoS.}
\linespread{1}\selectfont
\begin{tabular}{|p{2.1cm}<{\centering}|p{3cm}<{\centering}|}
\hline  \bf{Attack} & \bf{Ratio} \\ \hline
XSS & 0.695806 \\ \hline
Injection &  0.124408 \\ \hline
Password   &   0.124337 \\ \hline
Scanning   &   0.028964 \\ \hline
Backdoor   &   0.021716 \\ \hline
Ransomware  &  0.004078 \\ \hline
\end{tabular}
\label{tbl:attack_ratio_train_toniot}
\end{table}

The ToN\_IoT training subset is split into train/validation/test partitions (70:15:15). The universal model is trained on 1,748,571 samples. Unless noted otherwise, all reported results use the same feature-selection and preprocessing pipeline described in Section~\ref{sec:eval_umodel}.

\textbf{UNSW-NB15.} For cross-domain evaluation, we use the UNSW-NB15 dataset \cite{moustafa2015unswnb15}, which is commonly used for benchmarking network intrusion detection under different traffic characteristics. We treat UNSW-NB15 as a domain-shift target to examine how ZAID’s adapter-based localised model behaves when transferred beyond ToN\_IoT-style IoT traffic.

\subsection{Evaluation metrics}
ZAID’s performance is reported using accuracy, precision, recall, and F1 score (Table~\ref{tb:per_metric}). These metrics are reported because class imbalance and deployment shift can make accuracy alone misleading, particularly when rare but operationally important attack families are withheld during training.

\begin{table}[!t]
\footnotesize
\renewcommand\arraystretch{2}
\centering
\caption{Performance metrics.}
\linespread{1}\selectfont
\begin{tabular}{|p{1.1cm}<{\centering}|p{6cm}<{\centering}|}
\hline  \bf{Metric} & \bf{Equation} \\ \hline
Recall & $\dfrac{TP}{TP + FN}$ \\ \hline
Precision & $\dfrac{TP}{TP + FP}$ \\ \hline
Accuracy & $\dfrac{TP + TN} {TP + TN + FP + FN}$ \\ \hline
F1 Score & 2 $\times$ $\dfrac{precision \times recall}{precision + recall}$ \\ \hline
\end{tabular}
\label{tb:per_metric}
\vspace{1ex}
{\raggedright $^*$ True Positive (TP), False Negative (FN), False Positive (FP), and True Negative (TN). \par}
\end{table}

\subsection{Baseline selection and reporting policy}
\label{sec:eval_baselines_policy}
To mitigate potential concerns regarding baseline selection, we partition the comparisons into two groups:

\textbf{(A) Protocol-aligned baselines (main comparison).}
These studies are closest to ZAID’s core themes (federated collaboration and/or transfer/adaptation for intrusion detection) and can be meaningfully adapted to the same feature pipeline and zero-day protocol, such as Popoola \emph{et al.} \cite{popoola2021federated} as a federated deep learning for zero-day botnet detection at IoT-edge devices, and Bierbrauer \emph{et al.} \cite{bierbrauer2023transfer} for transfer-learning-centric intrusion detection framing.

\textbf{(B) Context-only reference points from prior work.}
We additionally include two implementations inspired by broader FL/security systems literature: Mohamed \emph{et al.} \cite{mohamed2023digital} and Khan \emph{et al.} \cite{rp11}. These works are thematically related (federated learning in IoT/security settings) but are not natively defined around ZAID’s deployment protocol (zone split, pseudo-labelling weighted anomaly score, and withheld-family zero-day evaluation on ToN\_IoT/UNSW-NB15). We therefore report them in a separate table and treat them as contextual reference points rather than headline baselines.

\subsection{Universal model and autoencoder evaluation}
\label{sec:eval_umodel}
\begin{figure*}[!t]
    \centering
    \includegraphics[keepaspectratio, scale=0.6]{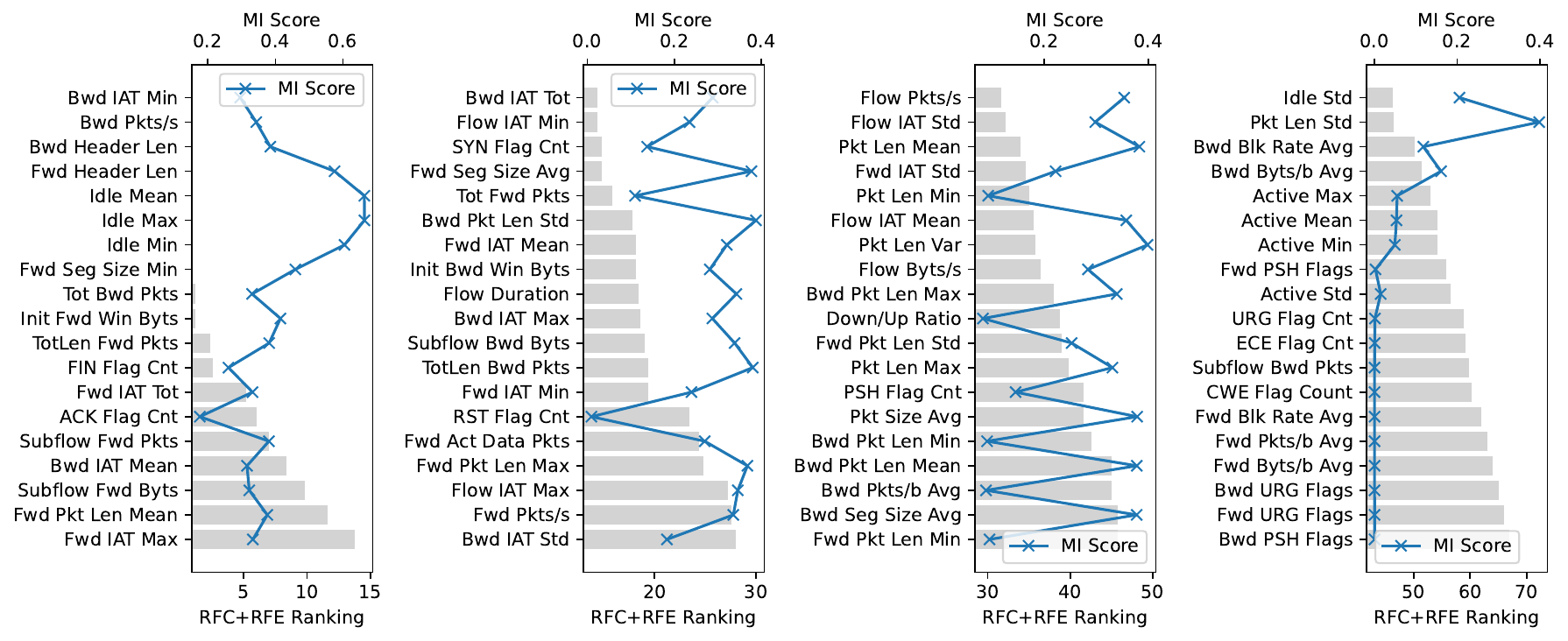}
    \caption{Feature selection ranking using RFC and RFE is on the bottom x-axis, while the mutual information score is on the top x-axis. Lower rankings and higher MI scores are preferable.}
    \vspace{1pt}
    \parbox{1\textwidth}{\small \textit{Note:} Forward (Fwd), Backward (Bwd), Length (Len), Segment (Seg), Total Length (TotLen), Total (Tot), Inter-arrival Time (IAT), Packet (Pkt), Average (Avg), Standard deviation (Std), Count (Cnt), Bulk (Blk).}
    \label{fig:feature_selection}
\end{figure*}

\textbf{Feature selection.}
We first perform feature selection on the ToN\_IoT training data using RFC+RFE alongside mutual information. Flow identifiers, source/destination IP addresses and ports, timestamps, and protocol identifiers are removed. Figure~\ref{fig:feature_selection} shows the ranked features and MI scores. We select features from the top 40\% of MI scores, resulting in 32 features.

\textbf{Training configuration.}
The ZAID universal model and autoencoder use a learning rate of $0.001$ and a batch size of 128 over 30 epochs, following the configuration used in \cite{mohamed2023digital}. We use model checkpointing based on the validation F1 score and validation loss to save the best model at each epoch.

\begin{figure}[!t]
  \centering
    \begin{minipage}[b]{0.45\columnwidth}
    \centering
    \includegraphics[keepaspectratio,scale=0.32]{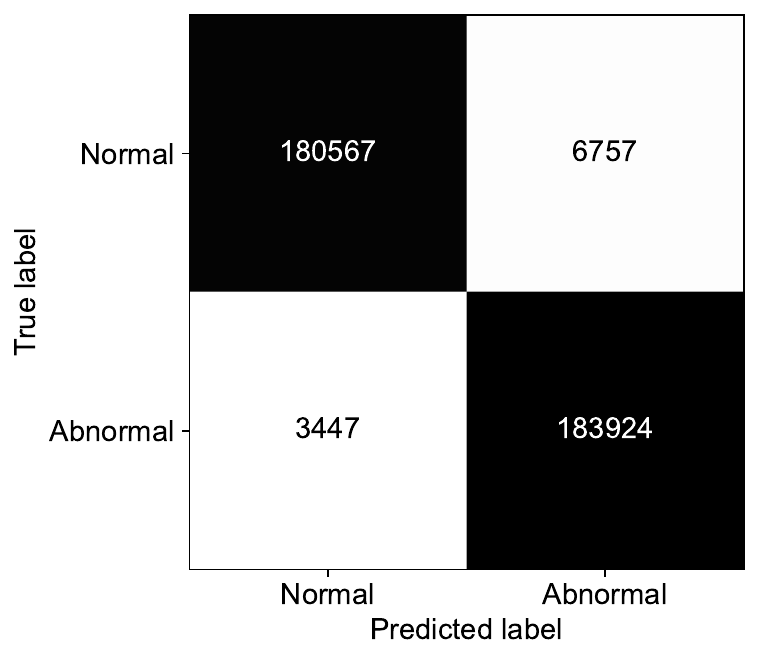}
    \subcaption{Confusion matrix}
    \label{fig:umodel_clra}
  \end{minipage}
  \begin{minipage}[b]{0.53\columnwidth}
  \centering
    \includegraphics[keepaspectratio,scale=0.42]{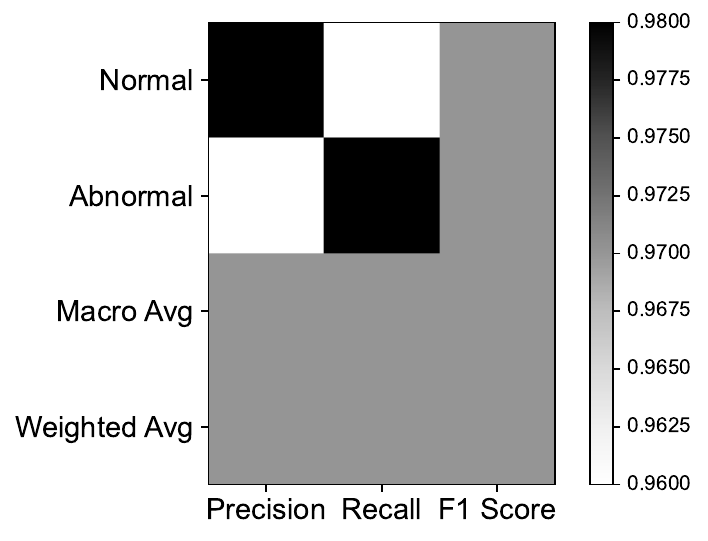}
    \subcaption{Classification report}
    \label{fig:umodel_clrb}
  \end{minipage}
    \begin{minipage}[b]{1\columnwidth}
  \centering
    \includegraphics[keepaspectratio,scale=0.42]{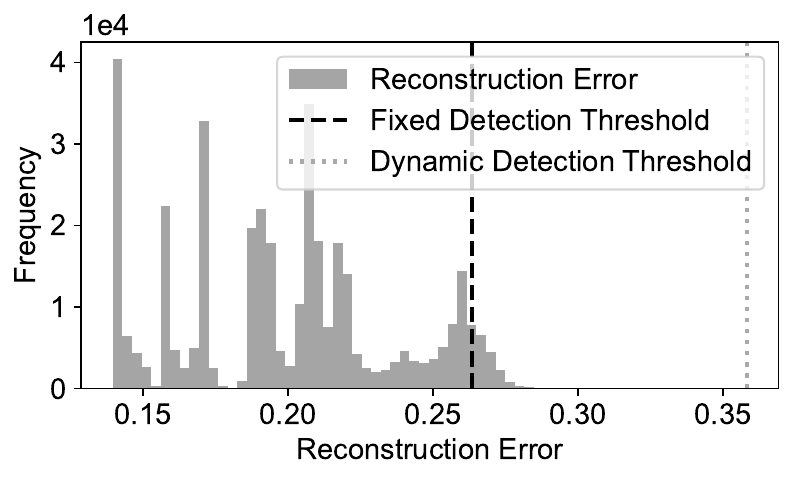}
    \subcaption{Autoencoder reconstruction error distribution.}
    \label{fig:umodel_clrc}
  \end{minipage}
 \caption{ZAID universal model (a) confusion matrix and (b) classification report. Macro average equally weights classes in the final metric. (c) Reconstruction error distribution of the autoencoder on test datasets.}
 \label{fig:umodel_cm}
\end{figure}

\textbf{Universal model performance.}
The ZAID universal model achieves an accuracy of 97.28\%, with 96.45\% precision and 98.16\% recall (Table~\ref{tbl:cid_static_result}). Figure~\ref{fig:umodel_cm} shows the confusion matrix and classification report.

\textbf{Autoencoder behaviour.}
The autoencoder uses the 95$^{\text{th}}$ percentile of reconstruction error as a detection threshold; samples exceeding it are flagged as anomalous/adversarial. This flags 18,735 samples out of 374,695 (i.e., the top 5\% by reconstruction error). Figure~\ref{fig:umodel_clrc} shows the reconstruction-error distribution.

However, percentile thresholding imposes a fixed alert rate and implicitly assumes that anomalies lie in the extreme tail of a stationary error distribution. In evolving environments (concept drift), reconstruction errors can shift and become multi-modal, and anomalous/adversarial samples may overlap with normal samples rather than forming a clean tail. To reduce sensitivity to these effects, we derive a dynamic threshold by clustering the normalised reconstruction errors (e.g., $k{=}2$ KMeans) and separating the low-error (nominal) and high-error (anomalous) groups. This adapts the decision boundary to the current error distribution instead of fixing it to a pre-specified percentile. 

Let $r = [r_1, r_2, ..., r_N]$ represent reconstruction errors, $r_i \in \mathbb{R_{\geq 0}}$. We normalise:
\begin{equation}
\tilde{r}_i = \frac{r_i - \min(\mathbf{r})}{\max(\mathbf{r}) - \min(\mathbf{r})}, \quad \forall i \in \{1, \dots, N\}.
\end{equation}
We then apply KMeans on $\tilde{\mathbf{r}}$ with $k=2$:
\begin{equation}
    \mathcal{C}_0, \mathcal{C}_1 = \text{KMeans}(\tilde{\mathbf{r}}, k=2).
\end{equation}
The anomalous cluster is the one with the higher centroid:
\begin{equation}
    \mathcal{C}_\text{anom} = \arg\max_{j \in \{0,1\}} \left( \mu_j \right), \quad
    \mu_j = \frac{1}{|\mathcal{C}_j|} \sum_{\tilde{r}_i \in \mathcal{C}_j} \tilde{r}_i.
\end{equation}
A dynamic threshold can be computed as:
\begin{equation}
    \tau = \min_{\tilde{r}_i \in \mathcal{C}_\text{anom}} \tilde{r}_i,
\end{equation}
and anomaly labels are assigned by:
\begin{equation}
    a_i =
\begin{cases}
1 & \text{if } \tilde{r}_i > \tau, \\
0 & \text{otherwise}.
\end{cases}
\end{equation}

\begin{table}[!h]
\footnotesize
\renewcommand\arraystretch{1.1}
\centering
\caption{Single-site performance metrics comparison on ToN\_IoT (universal model setting).}
\linespread{1}\selectfont
\begin{tabular}{|p{1.3cm}<{\centering}|p{1.2cm}<{\centering}|p{1.2cm}<{\centering}|p{1.2cm}<{\centering}|p{1.2cm}<{\centering}|}
\hline  \bf{Study} & \bf{Accuracy}  & \bf{Precision}  & \bf{Recall}  & \bf{F1 Score} \\ \hline
ZAID  & 97.28 & 96.45 & 98.16 & 97.27 \\ \hline
Bierbrauer et al. \cite{bierbrauer2023transfer} & 97.33 & 96.51 & 98.21 & 97.33 \\ \hline
Mohamed et al. \cite{mohamed2023digital} & 97.06 & 96.22 & 97.95 &  97.06 \\ \hline
Popoola et al. \cite{popoola2021federated} &  97.35 & 96.48 & 98.27 & 97.35\\ \hline
Khan et al. \cite{rp11} & 96.59 & 94.93 & 97.16 & 96.00 \\ \hline
\end{tabular}
\label{tbl:cid_static_result}
\end{table}

\subsection{Collaborative model preparation}
\label{sec:collaborative_preparation}
To prepare ZAID for zone deployment, we recycle the universal model to create a lightweight adapter-based localised model (Figure~\ref{fig:localized_dnna}). The universal model parameters remain frozen and identical across zones, while adapter parameters are trainable per zone.

We initialise the adapter-based model by training for one-third of the universal-model epochs (10) on a subset of 374,694 samples drawn from the unseen training data (excluding the three withheld zero-day attacks). This initialisation aims to align the adapter parameters with the base representation without duplicating capacity already captured in the frozen universal model.

We then construct four zone partitions (four military zones) using nearly 160,000 unseen network traffic records. Each partition includes the withheld attack families (MITM, DDoS, DoS) and is distributed evenly across zones. Table~\ref{tbl:attack_ratio_zone_toniot} reports the per-zone class counts.

\begin{table}[!b]
\footnotesize
\renewcommand\arraystretch{1.1}
\centering
\caption{Attack proportion in each zone based on ToN\_IoT.}
\linespread{1}\selectfont
\begin{tabular}{|p{1.5cm}<{\centering}|p{1.2cm}<{\centering}|p{1.2cm}<{\centering}|p{1.2cm}<{\centering}|p{1.2cm}<{\centering}|}
\hline  \bf{Attack} & \bf{$Z_1$} & \bf{$Z_2$} & \bf{$Z_3$} & \bf{$Z_4$} \\ \hline
Benign & 22,005 & 22,161 &22,313 & 22,294  \\ \hline
XSS & 18,077 & 17,970& 17,847 & 17,851  \\ \hline
Injection &  1,634 & 1,717&1,656 & 1,699  \\ \hline
Password   &   2,669 & 2,552 &2,595 & 2,556 \\ \hline
MITM      &    29 & 13 &12 & 13  \\ \hline
DDoS      &    8 & 7 &3 & 6 \\ \hline
DoS       &    4 & 6&- & 7  \\ \hline
\end{tabular}
\label{tbl:attack_ratio_zone_toniot}
\vspace{1ex}
{\raggedright $^*$ Total samples per zone is 39,983. \par}
\end{table}

To assign labels to unseen traffic at the zone level, we use the weighted anomaly score $AS_\omega$ (Equation~\ref{eq:anomaly_score}) with weights $(\omega_{\alpha}, \omega_{\gamma}, \omega_{\beta}) \in \{0,1,1.5\}$ and a threshold that requires combined contribution from at least two components when using summation thresholds (e.g., $\geq 2$ or $\geq 2.5$), consistent with the current implementation.


We simulate inter-zone connectivity constraints using publicly available information on Combat Service Support Very Small Aperture Terminal (CSS VSAT) links~\cite{vsat}. For GEO/GSO satellite backhaul, propagation delay alone yields round-trip times (RTTs) on the order of hundreds of milliseconds (approximately $500$\,ms), with additional delay introduced by terrestrial routing, processing, and queueing~\cite{allman1999rfc2488}. In our experiments, we adopt this GEO-like setting to emulate a constrained backhaul for transferring model parameters between zones, rather than raw traffic. While modern Low Earth Orbit (LEO) constellations, e.g., Starlink, can provide substantially lower RTTs, our choice should be interpreted as a conservative, degraded-connectivity assumption consistent with legacy VSAT deployments.

\subsection{ZAID collaborative improvement via federated learning}
Each deployed zone trains its model on local traffic and periodically shares adapter parameters with a federated server for aggregation. In our setup, the collaborative model is retrained locally for 2 epochs (learning rate 0.0005) and participates in five aggregation rounds. To tolerate zone or network failures, if a zone does not receive the updated global model (timeout of 50 seconds), it falls back to the latest successfully received global model; otherwise, the initially deployed model remains active. If all zones fail in a round, the latest server-stored model is used.

\textbf{Protocol-aligned comparisons (headline results).}
Table~\ref{tbl:final_performance_core_toniot} reports the average performance across aggregation rounds, focusing on protocol-aligned baselines and ZAID variants. ZAID achieves its best average accuracy of \textbf{83.16\%} on the withheld-family (zero-day) ToN\_IoT setting using \textbf{ZAID (1, 1.5, 1.5) $\geq$ 2.5}.

\begin{table}[!t]
\footnotesize
\renewcommand\arraystretch{1.1}
\centering
\caption{Average performance metrics comparison on ToN\_IoT under the withheld-family zero-day protocol (protocol-aligned comparisons).}
\linespread{1}\selectfont
\begin{tabular}{|p{2.8cm}<{\centering}|p{1cm}<{\centering}|p{1cm}<{\centering}|p{1cm}<{\centering}|p{1cm}<{\centering}|}
\hline  \bf{Study} & \bf{Accuracy} & \bf{Precision} & \bf{Recall} & \bf{F1 Score} \\ \hline
Bierbrauer et al. \cite{bierbrauer2023transfer} & 58.39 & 34.55& 77.71 & 23.23  \\ \hline
Popoola et al. \cite{popoola2021federated} &  39.52 & 33.71&37.71 & 31.81  \\ \hline
ZAID (1, 0, 0) $\geq$ 1   &  64.27 & 80.76 & 36.27 & 47.66 \\ \hline
ZAID (0, 1, 0) $\geq$ 1   &  82.34 & 97.59 & 66.24 & 78.21 \\ \hline
ZAID (0, 0, 1) $\geq$ 1   &  81.80 & 81.37 & \textbf{86.14} & \textbf{81.95} \\ \hline
ZAID (1, 1, 1) $\geq$ 2  &  81.57 &  97.70 & 64.59 & 77.07\\ \hline
ZAID (1, 1.5, 1) $\geq$ 2.5   &  82.50 &  97.77 & 66.45 & 78.42\\ \hline
ZAID (1.5, 1.5, 1) $\geq$ 2.5   &   82.25 &97.70 & 66.00 & 78.08 \\ \hline
\textbf{ZAID (1, 1.5, 1.5) $\geq$ 2.5}   &   \textbf{83.16} & 97.69 & 67.86 & 79.40 \\ \hline
ZAID (1.5, 1, 1.5) $\geq$ 2.5   &  82.86 & \textbf{97.85} & 67.04 & 78.88\\ \hline
\end{tabular}
\label{tbl:final_performance_core_toniot}
\end{table}

\textbf{Reference-inspired baselines (context-only).}
For completeness and transparency, Table~\ref{tbl:final_performance_ref_toniot} reports additional results for two reference-inspired baselines. These are included as contextual points of comparison but are not treated as headline baselines for ZAID due to differences in original problem framing and deployment protocol.

\begin{table}[!t]
\footnotesize
\renewcommand\arraystretch{1.1}
\centering
\caption{Average performance on ToN\_IoT under the withheld-family zero-day protocol (reference-inspired baselines reported for context).}
\linespread{1}\selectfont
\begin{tabular}{|p{2.8cm}<{\centering}|p{1cm}<{\centering}|p{1cm}<{\centering}|p{1cm}<{\centering}|p{1cm}<{\centering}|}
\hline  \bf{Study} & \bf{Accuracy} & \bf{Precision} & \bf{Recall} & \bf{F1 Score} \\ \hline
Mohamed et al. \cite{mohamed2023digital} & 50.58 & 2.07 & 16.43 & 1.11  \\ \hline
Khan et al. \cite{rp11} & 49.55 & 98.35 & 49.76 & 65.62  \\ \hline
\end{tabular}
\label{tbl:final_performance_ref_toniot}
\end{table}

\subsection{Cross-domain evaluation with UNSW-NB15}
\label{sec:eval_crossdomain}
We investigate whether ZAID’s adapter-based collaborative model can transfer under domain shift from ToN\_IoT-style IoT traffic to UNSW-NB15 enterprise-style network traffic \cite{moustafa2015unswnb15}. We reuse the pretrained collaborative model from Section~\ref{sec:collaborative_preparation} and adapt it using the same deployment procedure. This reflects the practical constraint that fully fine-tuning the universal model can be computationally expensive and may risk forgetting previously learned representations; adapter updates offer a lightweight alternative.

In \textsc{UNSW-NB15}, we treat the rarer attack categories as withheld (zero-day) classes for evaluation and retrain on the remaining categories under the same protocol logic. Specifically, the withheld categories are \emph{Analysis}, \emph{Backdoor}, \emph{DoS}, \emph{Generic}, \emph{Shellcode}, and \emph{Worms}. We retrain the relevant models using the training subset containing \emph{Exploits} (27{,}856), \emph{Fuzzers} (26{,}652), \emph{Reconnaissance} (15{,}061), and \emph{Benign} (41{,}225), while preserving the same cross-domain evaluation procedure. Table~\ref{tbl:attack_ratio_augment} reports the resulting per-zone class counts used in the current implementation.

\textbf{Protocol-aligned comparisons.}
Table~\ref{tbl:final_performance_core_unsw} reports cross-domain results for protocol-aligned baselines and \textsc{ZAID} variants. The \textsc{ZAID} collaborative model achieves the best accuracy of \textbf{71.64\%} under \textbf{\textsc{ZAID} (0, 1, 0) $\geq$ 1} in this domain-shift setting, indicating stronger generalisation relative to the protocol-aligned baselines. Notably, although auxiliary components (e.g., the autoencoder/universal models, where applicable) are not re-tuned on \textsc{UNSW-NB15} in this experiment, their performance remains close to several baseline studies, suggesting partial robustness under cross-domain transfer. This also motivates fine-tuning the full pipeline when adapting to new datasets to improve sensitivity to newly observed traffic patterns.


\begin{table}[!h]
\footnotesize
\renewcommand\arraystretch{1.1}
\centering
\caption{Attack proportion in each zone based on ToN\_IoT \& UNSW-NB15.}
\linespread{1}\selectfont
\begin{tabular}{|p{1.7cm}<{\centering}|p{1.2cm}<{\centering}|p{1.2cm}<{\centering}|p{1.2cm}<{\centering}|p{1.2cm}<{\centering}|}
\hline  \bf{Attack} & \bf{$Z_1$} & \bf{$Z_2$} & \bf{$Z_3$} & \bf{$Z_4$} \\ \hline
Benign & 23,162 & 23,261 &23,371 & 23,397  \\ \hline
XSS & 18,031 & 18,001& 17,834 & 17,836  \\ \hline
Injection &  1,696 & 1,678 & 1,727 & 1,673  \\ \hline
Password   &   2,606 & 2,548 &2,553 & 2,583 \\ \hline
Analysis   &   5 & 9 & 13 & 16 \\ \hline
Backdoor   &   14 & 15 & 13 & 9 \\ \hline
Exploits  &  86 & 98 & 93 & 100\\ \hline
MITM      &    16 & 14 &11 & 17  \\ \hline
DDoS      &    5 & 7 &7 & 5 \\ \hline
DoS       &    121 & 137& 157 & 134  \\ \hline
Fuzzers & 95 & 97 & 74 & 93\\ \hline
Reconnaissance & 61 & 46 &  42 & 46\\ \hline
Shellcode & 69 & 69 & 57 & 62 \\ \hline
Worms &7 &5 & 9 &4\\ \hline
Generic & 133 & 125 & 147& 135\\ \hline
\end{tabular}
\label{tbl:attack_ratio_augment}
\vspace{1ex}
{\raggedright $^*$ Total samples per zone is 41,497. \par}
\end{table}

\begin{table}[!t]
\footnotesize
\renewcommand\arraystretch{1.1}
\centering
\caption{Average performance metrics comparison under domain shift to UNSW-NB15 (protocol-aligned comparisons).}
\linespread{1}\selectfont
\begin{tabular}{|p{2.8cm}<{\centering}|p{1cm}<{\centering}|p{1cm}<{\centering}|p{1cm}<{\centering}|p{1cm}<{\centering}|}
\hline  \bf{Study} & \bf{Accuracy} & \bf{Precision} & \bf{Recall} & \bf{F1 Score} \\ \hline
Bierbrauer et al. \cite{bierbrauer2023transfer} & 64.90 & 63.71& \textbf{68.02} & 65.79  \\ \hline
Popoola et al. \cite{popoola2021federated} &  66.74 & 73.39&51.77 & 59.82  \\ \hline
ZAID (1, 0, 0) $\geq$ 1   &  62.52 & 75.87 & 35.81 & 46.35 \\ \hline
\textbf{ZAID (0, 1, 0) $\geq$ 1}   &  \textbf{71.64} & 75.99 & 62.75 & \textbf{67.94} \\ \hline
ZAID (0, 0, 1) $\geq$ 1  &  62.03 & 76.06 & 35.27 &44.55 \\ \hline
ZAID (1, 1, 1) $\geq$ 2   &  65.21 &  76.88 & 43.14 & 53.46\\ \hline
ZAID (1, 1.5, 1) $\geq$ 2.5   &  61.01 &  74.39 & 32.41 & 42.24\\ \hline
ZAID (1.5, 1.5, 1) $\geq$ 2.5 &   66.16 & \textbf{77.14} & 45.47 & 55.67 \\ \hline
ZAID (1, 1.5, 1.5) $\geq$ 2.5  &   65.09 & 76.80 & 42.76 & 53.17 \\ \hline
ZAID (1.5, 1, 1.5) $\geq$ 2.5  &  64.57 & 76.90 & 41.15 & 51.73\\ \hline
\end{tabular}
\label{tbl:final_performance_core_unsw}
\end{table}

\textbf{Reference-inspired baselines (context-only).}
For transparency, Table~\ref{tbl:final_performance_ref_unsw} reports reference-inspired baselines under the same cross-domain procedure; however, we avoid treating them as headline comparisons for the same reasons discussed in Section~\ref{sec:eval_baselines_policy}.

\begin{table}[!t]
\footnotesize
\renewcommand\arraystretch{1.1}
\centering
\caption{Average performance under domain shift to UNSW-NB15 (reference-inspired baselines reported for context).}
\linespread{1}\selectfont
\begin{tabular}{|p{2.8cm}<{\centering}|p{1cm}<{\centering}|p{1cm}<{\centering}|p{1cm}<{\centering}|p{1cm}<{\centering}|}
\hline  \bf{Study} & \bf{Accuracy} & \bf{Precision} & \bf{Recall} & \bf{F1 Score} \\ \hline
Mohamed et al. \cite{mohamed2023digital} & 66.07 & 76.01 & 46.34 & 56.57  \\ \hline
Khan et al. \cite{rp11} & 65.38 & 69.69 & 54.34 & 59.73  \\ \hline
\end{tabular}
\label{tbl:final_performance_ref_unsw}
\end{table}

\subsection{Limitations and threats to validity}
\label{sec:limitations}
\textbf{Dataset and scenario representativeness.}
Our evaluation relies on ToN\_IoT and UNSW-NB15, which are common in intrusion detection but do not fully capture the diversity and constraints of real IoBT deployments, including mission dynamics, radio artefacts, and adversarial disruption. The experiment 'zones' are simulated dataset partitions that approximate rather than replicate tactical zone dynamics.

\textbf{Definition of ``zero-day''.}
We define zero-day conditions by withholding specific attack types during supervised training and introducing them later during zone adaptation. This tests unseen attack categories but does not cover all real-world novelty forms, such as subtle variations, multi-stage campaigns, or concept drift where normal behaviour evolves. Future work should explore more models of novelty and drift.

\textbf{Pseudo-labelling noise and threshold sensitivity.}
ZAID uses pseudo-labelling to analyse weakly labelled zone traffic and an autoencoder signal as an auxiliary indicator. Inaccurate confidence calibration can cause systematic errors, and shifts in traffic can affect reconstruction-error thresholding. We present results with the current dynamic threshold; future work could explore adaptive thresholding and uncertainty-aware pseudo-labelling.

\textbf{Federated learning assumptions and adversarial robustness.}
Our federated setup follows standard aggregation, assuming zones act honestly when contributing updates. In practice, updates can be manipulated (e.g., poisoning or backdoor attacks) or leak local data through gradients or updates. Defences such as robust aggregation, validation, secure aggregation, and privacy mechanisms are important but beyond this paper's scope.

\textbf{Comparative evaluation and reproducibility.}
Several external works address related themes (e.g., federated learning in IoT security or cyber defence in other domains) but do not align with ZAID's deployment protocol (zone split, withheld-family evaluation, and weighted pseudo-labelling with anomaly signal). To minimise comparability risk, we distinguish protocol-aligned comparisons from contextual references. Reproducibility depends on matching preprocessing and training budgets; thus, we emphasise ZAID ablations and protocol-aligned baselines as the most reliable comparisons.

\textbf{Communication and systems effects.}
ZAID is motivated by limited communications, focusing on detection performance rather than a systems-level analysis of bandwidth, energy, or latency. While ZAID avoids transmitting raw traffic by exchanging model parameters, understanding its full impact requires dedicated studies and deployment measurements.

\section{Conclusion}
\label{sec:conclusion}
This paper presented \textbf{Z}one-\textbf{A}daptive \textbf{I}ntrusion \textbf{D}etection (\textbf{ZAID}), a zone-centric framework inspired by IoBT environments characterised by intermittent connectivity, limited bandwidth, and variable traffic across zones. ZAID utilises three mechanisms to ensure robustness against distribution shifts and new attack types: (a) a universal CNN trained on labelled data for shared detection, (b) an autoencoder using reconstruction error as an anomaly indicator, and (c) a parameter-efficient, adapter-based local model that updates at the zone level and improves via federated learning. 
We evaluated ZAID with ToN\_IoT under a withheld-family protocol, excluding attack categories (MITM, DDoS, DoS) from supervised training and introducing them during zone-level adaptation. The best ZAID achieved 83.16\% accuracy for collaborative detection. We also tested cross-domain transfer to UNSW-NB15, achieving 71.64\% accuracy. These results demonstrate the value of zone-personalised collaboration with lightweight parameters, avoiding raw traffic transmission.

\bibliographystyle{elsarticle-num}
\bibliography{main}







\end{document}